\documentclass[twocolumn,floatfix,a4paper,showpacs,showkeys,nofootinbib,reprint,prl]{revtex4-1}
\usepackage{epsfig}
\usepackage{latexsym}
\usepackage{xspace}
\usepackage[colorlinks=true,linktocpage=true,linkcolor=blue,citecolor=blue,allcolors=blue]{hyperref}
\usepackage{url}
\usepackage[utf8]{inputenc}
\usepackage{indentfirst}
\usepackage{enumerate}
\usepackage{color}

\usepackage[caption=false,position=top]{subfig}

\usepackage{amsmath}
\usepackage{amssymb}

\usepackage[english]{babel}
\usepackage{url}

\AtBeginDocument{\renewcommand{\natexlab}[1]{#1}}

\newcommand{\eq}[1]{\begin{align} #1 \end{align}}

\begin{document}

\title{
Quarkyonic or baryquark matter?\\ 
On the dynamical generation of momentum space shell structure
}

\author{Volker Koch}
\affiliation{Nuclear Science Division, Lawrence Berkeley National Laboratory, 1 Cyclotron Road, Berkeley, CA 94720, USA}

\author{Volodymyr Vovchenko}
\affiliation{Institute for Nuclear Theory, University of Washington, Box 351550, Seattle, WA 98195, USA}
\affiliation{Physics Department, University of Houston, Box 351550, Houston, TX 77204, USA}
\affiliation{Frankfurt Institute for Advanced Studies, Giersch Science Center,
D-60438 Frankfurt am Main, Germany}

\begin{abstract}
We study the equation of state of a mixture of (quasi-)free constituent quarks and nucleons with hard-core repulsion at zero temperature.
Two opposite scenarios for the realization of the Pauli exclusion principle are considered: 
(i) a Fermi sea of quarks surrounded by a shell of baryons -- the quarkyonic matter, and 
(ii) a Fermi sea of nucleons surrounded by a shell of quarks which we call \emph{baryquark matter}.
In both scenarios, the sizes of the Fermi sea and shell are fixed through energy minimization at fixed baryon number density.
While both cases yield a qualitatively similar transition from hadronic to quark matter,
we find that baryquark matter is energetically favored in this setup and yields a physically acceptable behavior of the speed of sound without the need to introduce an infrared regulator.
In order to retain the theoretically more appealing quarkyonic matter as the preferred form of dense QCD matter will thus require modifications to the existing dynamical generation mechanisms, such as, for example, the introduction of momentum-dependent nuclear interactions.
\end{abstract}

\maketitle

\paragraph{\bf Introduction.}

\begin{figure}[t]
  \centering
  \includegraphics[width=.49\textwidth]{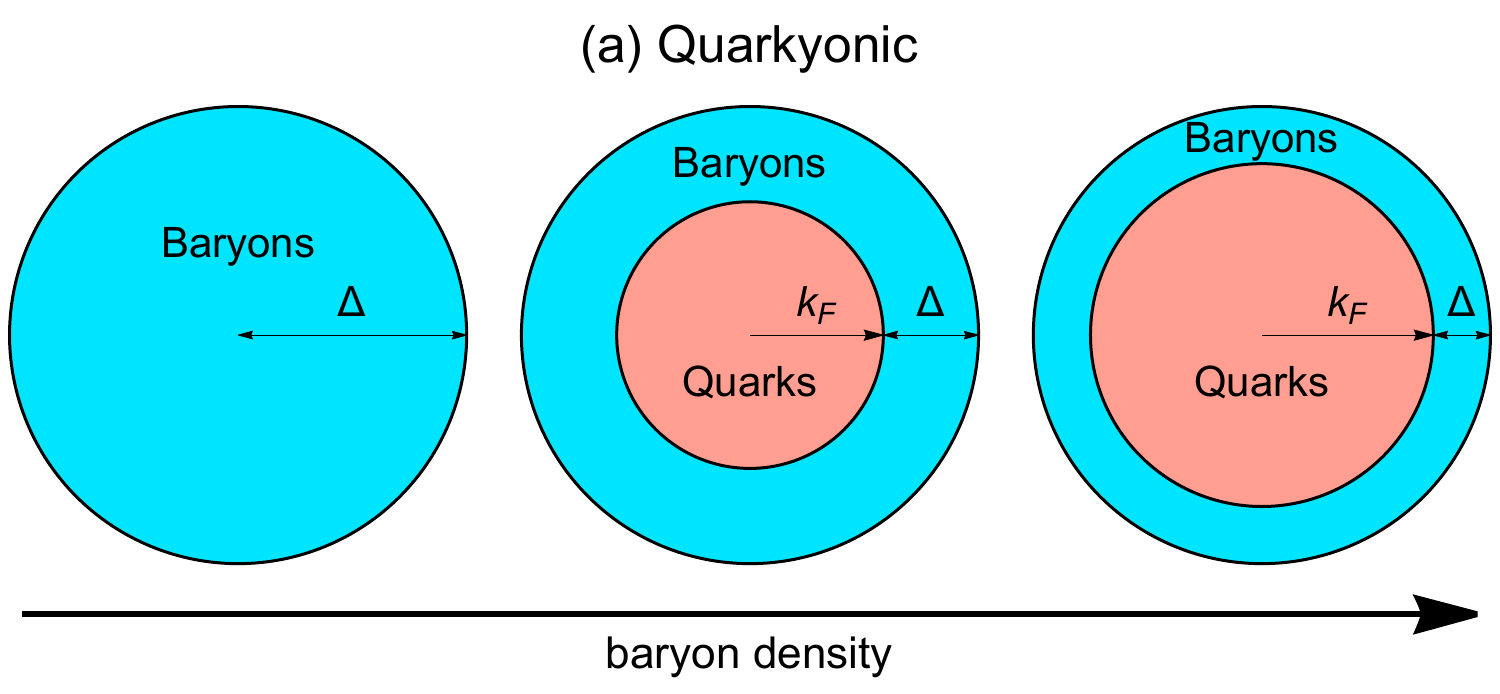}
  \vskip8pt
  \includegraphics[width=.49\textwidth]{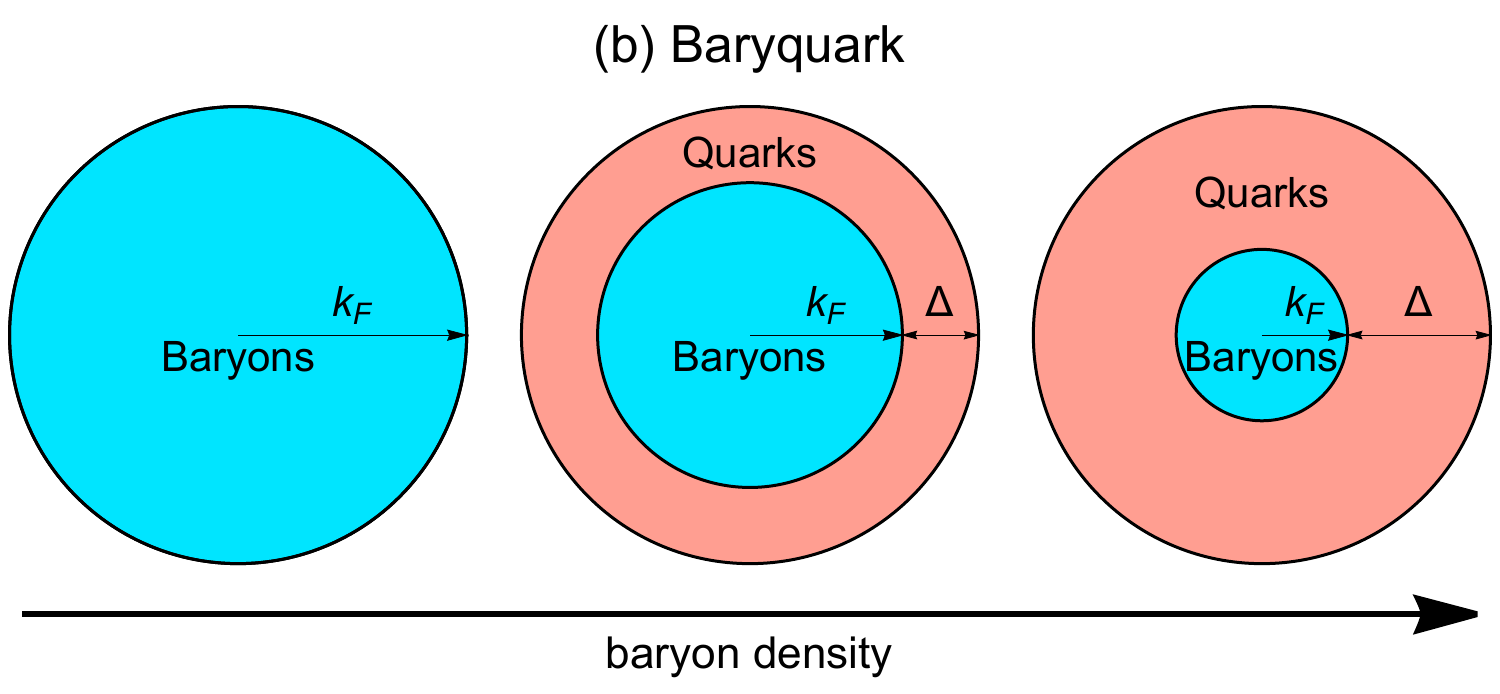}
  \caption{
  Evolution of the momentum shell structure in isospin symmetric (a) quarkyonic and (b) baryquark matter as a function of baryon number density at zero temperature.
  The momentum scales $k_F$ and $\Delta$ are given for baryon degrees of freedom, i.e. for quarks they should be divided by $N_c$.
  }
  \label{fig:shell}
\end{figure}

The concept of quarkyonic matter~\cite{McLerran:2007qj}, inspired by the expected QCD properties in the large $N_c$ limit, is a realization of quark-hadron duality where both degrees of freedom appear as quasiparticles.
It envisions the existence of a quark-baryon mixture with a mixed phase in momentum space which is consistent with the Pauli exclusion principle.
The hallmark feature of quarkyonic matter is that
excitations around the Fermi surface are baryonic, which could be realized by imposing a shell structure in the momentum space: 
a Fermi sea of quasi-free, ``deconfined'' quarks surrounded by a shell of confined baryons~(Fig.~\ref{fig:shell}a).

One can imagine how the momentum shell structure emerges dynamically~\cite{McLerran:2018hbz}.
At small baryon densities, the matter is purely baryonic, reflecting that, at fixed baryon density, the energy per baryon of a free gas of nucleons is smaller than that of constituent quarks.
As the baryon density is increased, the hard-core repulsive interactions between nucleons become important and make the appearance of a quark Fermi sea more favorable relative to pure nucleon matter~\cite{Jeong:2019lhv}.
The matter consists primarily of quarks at large baryon densities, with an increasingly thin baryon shell~\cite{McLerran:2018hbz,Jeong:2019lhv}.
The resulting equation of state exhibits a soft-hard evolution, with the speed of sound containing a peak exceeding the conformal limit, in line with neutron star phenomenology~\cite{Tews:2018kmu,Fujimoto:2019hxv,Tan:2020ics,Altiparmak:2022bke}.
Various extensions and applications of this picture were studied in recent years~\cite{Han:2019bub,Sen:2020peq,Duarte:2020xsp,Zhao:2020dvu,Cao:2020byn,Duarte:2020kvi,Fukushima:2020cmk,Sen:2020qcd,Duarte:2021tsx}, as well as how such state of matter may emerge~\cite{Philipsen:2019qqm,Kovensky:2020xif,Kojo:2021ugu}.

However, it remains an open question whether quarkyonic matter is the energetically preferred form of matter. The dynamical generation of the momentum space structure considered in Refs.~\cite{McLerran:2018hbz,Jeong:2019lhv} has been performed under the assumption that a baryon Fermi shell on top of a quark Fermi sea is the only possible realization of the Pauli exclusion principle.
The Pauli principle alone, however, also permits other momentum space configurations.  
In this work, we confront the quarkyonic matter momentum space configuration with the opposite scenario we will call \emph{baryquark matter}, where the Fermi sea is filled with confined baryons surrounded by a Fermi shell of deconfined quarks.
We show that energy minimization of a baryon-quark mixture favors baryquark matter over quarkyonic matter although the resulting equations of state are qualitatively similar in both scenarios.
Some basic properties of the baryquark matter equation of state are explored. In particular, the model yields a physically acceptable behavior of the speed of sound without the need to introduce an infrared regulator.

\paragraph{\bf Baryon-quark mixture.}

We consider isospin symmetric nuclear matter at zero temperature and finite baryon density as a mixture of (quasi-)free quarks and nucleons. Along the lines of~\cite{McLerran:2018hbz}, we assume that chiral symmetry is broken to set the quark mass $m_Q = m_N / N_c$ as in the constituent quark model with no internal motion.
The hard-core repulsion among nucleons is incorporated via the excluded volume prescription \`a la van der Waals, where the system volume is substituted by the available volume $V \to V (1-n_N/n_0)$~\cite{Rischke:1991ke,Vovchenko:2015vxa}. Here $n_N$ is the nucleon number density, and $n_0 \equiv 1/b$ is the limiting density.
The baryon and energy densities consist of nucleon and quark contributions
\eq{
n_B &= n_N + n_Q, \\
\varepsilon &= \varepsilon_N + \varepsilon_Q.
}

The nucleon and quark contributions to $n_B$ and $\varepsilon$ depend on momentum space configurations for nucleons and quarks, which should respect the Pauli exclusion principle.
Due to the coinciding spin-isospin degeneracies of nucleons and quarks, $d_N = d_Q = 4$, the momentum levels occupied by deconfined quarks are unavailable to confined quarks, and vice versa.
The equilibrium configuration at fixed baryon density $n_B$ is obtained through the minimization of energy $\varepsilon$.
In the following, we consider quarkyonic and baryquark configurations.

\paragraph{\bf Quarkyonic matter.}

In the quarkyonic scenario, the quarks occupy the Fermi sea up to the Fermi momentum $k_F^q = k_F/N_c$ surrounded by a shell of nucleons with momenta between $k_F$ and $k_F + \Delta$.
The quark and nucleon densities read
\eq{
\label{eq:nQQM}
n_Q &= \frac{2}{\pi^2} \, \int_0^{k_F/N_c} \, dk \, k^2 = \frac{2 \, k_F^3}{3 \pi \, N_c^3},\\
n_N &= \frac{n_N^*}{1+n_N^*/n_0},
}
with
\eq{
n_N^* & = \frac{2}{\pi^2} \, \int_{k_F}^{k_F+\Delta} \, dk \, k^2 = \frac{2 \, [(k_F + \Delta)^3 - k_F^3]}{3\pi}.
}

The corresponding energy density contributions are
\eq{
\label{eq:eQQM}
\varepsilon_Q &= \frac{2 N_c}{\pi^2} \, \int_0^{k_F/N_c} \, dk \, k^2 \, \sqrt{m_Q^2 + k^2},\\
\varepsilon_N &= f_{\rm ev} \, \frac{2}{\pi^2} \, \int_{k_F}^{k_F+\Delta} \, dk \, k^2 \, \sqrt{m_N^2 + k^2}.
}
Here $f_{\rm ev} = (1+n_N^*/n_0)^{-1} = (1-n_N/n_0)$ is the excluded volume correction factor.

\paragraph{\bf Baryquark matter.}

In the baryquark scenario, the baryons occupy the Fermi sea up to the Fermi momentum $k_F$ surrounded by a shell of quarks with momenta between $k_F/N_c$ and $(k_F + \Delta)/N_c$.
Therefore,
\eq{
\label{eq:nQBQ}
n_Q &= \frac{2}{\pi^2} \, \int_{k_F/N_c}^{(k_F+\Delta)/N_c} \, dk \, k^2 = \frac{2 \, [(k_F+\Delta)^3 - k_F^3]}{3\pi \, N_c^3},\\
n_N &= f_{\rm ev} \, \frac{2}{\pi^2} \, \int_{0}^{k_F} \, dk \, k^2 = f_{\rm ev} \, \frac{2 \, k_F^3}{3\pi}.
}
and
\eq{
\label{eq:eQBQ}
\varepsilon_Q &= \frac{2 N_c}{\pi^2} \, \int_{k_F/N_c}^{(k_F+\Delta)/N_c} \, dk \, k^2 \, \sqrt{m_Q^2 + k^2},\\
\varepsilon_N &= f_{\rm ev} \, \frac{2}{\pi^2} \, \int_{0}^{k_F} \, dk \, k^2 \, \sqrt{m_N^2 + k^2}.
}

\paragraph{\bf Energy minimization.}

In quarkyonic and baryquark matter, the energy and baryon densities are defined through two momentum shell structure parameters, $k_F$ and $\Delta$. 
In particular, the values of $k_F$ and $\Delta$ regulate the quark fraction $n_Q/n_B$ at fixed baryon density $n_B$.
The equilibrium configuration for $k_F(n_B)$ and $\Delta(n_B)$ at given $n_B$ is thus obtained by minimizing the energy density $\varepsilon$ with respect to the quark fraction $n_Q/n_B$.
Performing the minimization for both the quarkyonic and baryquark configurations at the same baryon density allows one to establish the energetically preferred scenario.
This procedure was first introduced in Ref.~\cite{Jeong:2019lhv} and it differs from~\cite{McLerran:2018hbz} where a functional form for $\Delta$ was assumed instead.

To illustrate the procedure, we look at the dependence of the energy density $\varepsilon$ on the quark fraction in both scenarios at baryon density of $n_B = 4.8 \rho_0$.
We use the physical values of the parameters in our calculations, $N_c = 3$, $m_N = 0.938$~GeV/$c^2$, and $\rho_0 = 0.16$~fm$^{-3}$.

\begin{figure}[t]
  \centering
  \vskip4pt
  \includegraphics[width=.49\textwidth]{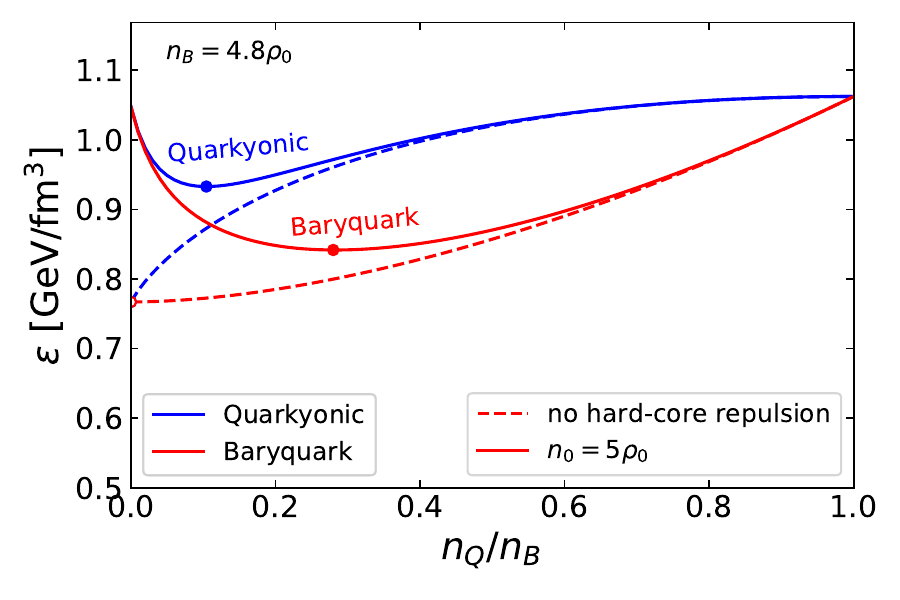}
  \caption{
  Dependence of the energy density in the excluded volume (blue) quarkyonic and (red) baryquark matter on the quark fraction $n_Q/n_B$ at a fixed baryon density of $n_B = 4.8 \rho_0$.
  The limiting nucleon density due to excluded volume in both cases is $n_0 \equiv b^{-1} = 5 \rho_0$.
  }
  \label{fig:emin}
\end{figure}

The blue and red lines in Fig.~\ref{fig:emin} depict the dependence of the energy density $\varepsilon$ on the quark fraction $n_Q/n_B$ in quarkyonic and baryquark scenarios, respectively.
First, we look at the results without the excluded volume correction, i.e. $n_0 \to \infty$ (dashed lines). 
In this case, it is always energetically favorable to have pure nucleon matter, $n_Q/n_B = 0$, at all baryon densities.
Already in the free gas limit, it is seen that for any non-zero quark fraction, the baryquark matter is energetically favored over the quarkyonic matter for all $0 < n_Q/n_B < 1$.

When the hard-core nucleon repulsion is included (solid lines), the energy minimum may correspond to a non-zero quark fraction. 
In this way, as first shown in Ref.~\cite{Jeong:2019lhv} for quarkyonic matter, the deconfined quarks emerge dynamically.
For $n_0 = 5 \rho_0$ that we take here, the energy density exhibits a minimum at $n_Q/n_B > 0$ at $n_B = 4.8 \rho_0$, both for quarkyonic and baryquark configurations~(solid points in Fig.~\ref{fig:emin}).
As in the case of non-interacting nucleons, the baryquark shell structure is energetically favored over the quarkyonic one.

\begin{figure*}[t]
  \centering
  \vskip4pt
  \includegraphics[width=\textwidth]{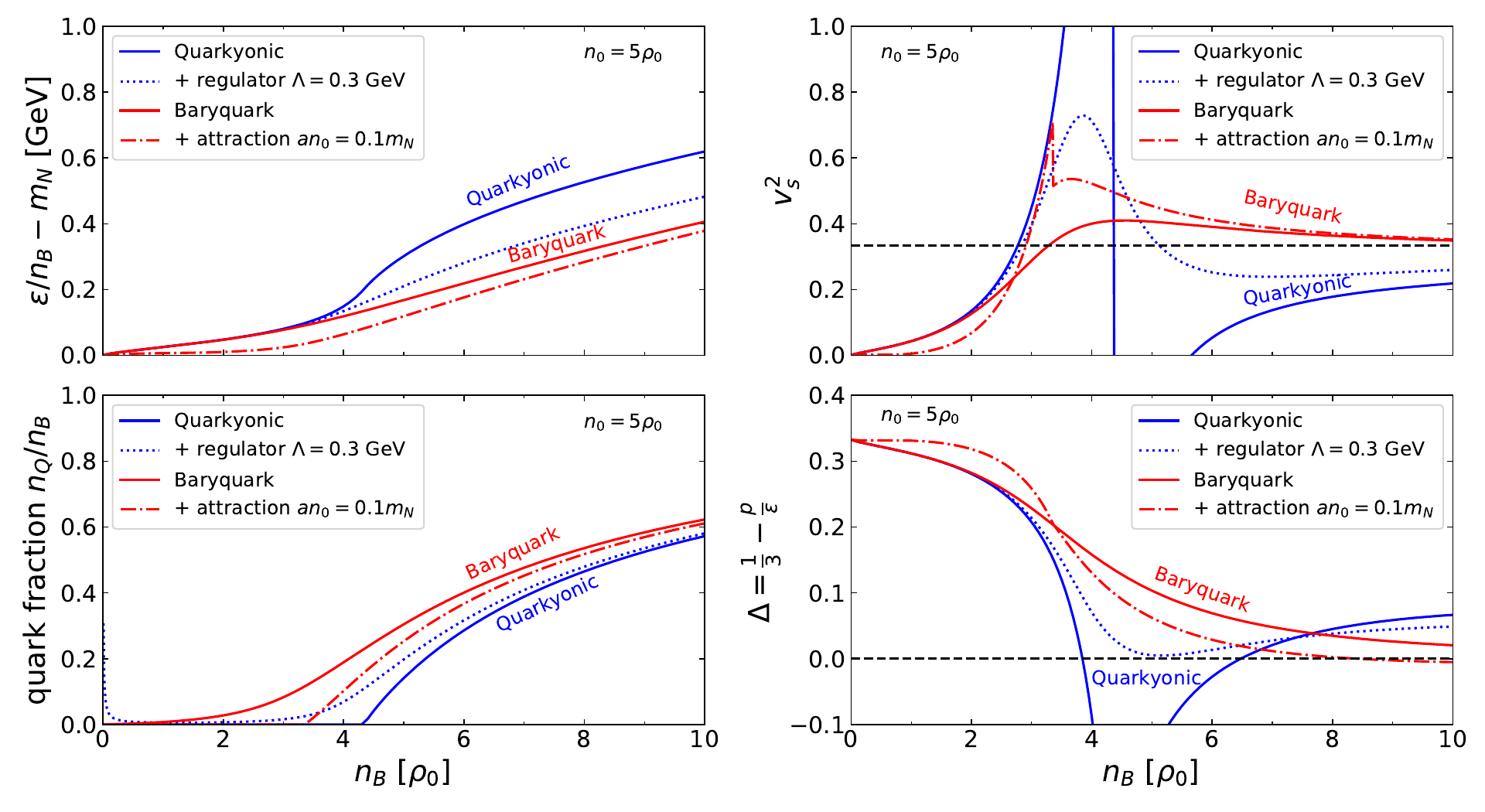}
  \caption{
  Baryon density dependence of energy per baryon (upper left), quark fraction (lower left), speed of sound squared (upper right), and trace anomaly (lower right) evaluated in excluded volume quarkyonic (blue lines) and baryquark~(red lines) matter.
  The excluded volume parameter in all cases corresponds to the limiting nucleon density of $n_0 \equiv b^{-1} = 5 \rho_0$.
  The dotted blue lines correspond to quarkyonic matter with an infrared regulator $\Lambda = 0.3$~GeV while the dash-dotted red lines depict baryquark matter with attractive nucleon mean-field, $a n_0 = 0.1 m_N$.
  }
  \label{fig:EoS}
\end{figure*}

\paragraph{\bf Equation of state.}

We now turn to the resulting equation of state.
The upper left panel of Fig.~\ref{fig:EoS} depicts the baryon density dependence of the energy per baryon.
To make the differences between the different scenarios more visible, we subtract the nucleon mass, $m_N$. 
Our results indicate that baryquark matter is energetically favored over quarkyonic matter at all baryon densities.
The equations of state in the two scenarios are virtually identical at low densities, $n_B \lesssim 3 \rho_0$, where the nucleons dominate.
Differences become visible at larger densities, where the equation of state of baryquark matter is notably softer than the quarkyonic one.
The difference is caused by an earlier appearance of quarks in the former case, as shown by the quark fraction in the lower left panel of Fig.~\ref{fig:EoS}.
In fact, the gradual appearance of quarks in the baryquark matter occurs already at low densities, although the quark fraction stays small until $n_B \simeq 2 \rho_0$.
As we discuss below, the appearance of quarks is shifted to higher densities once attractive nucleon interactions at low densities are taken into account.
In the quarkyonic scenario, the quarks only appear above a threshold density of $n_B \approx 4.4 \rho_0$.

The speed of sound $v_s^2 = dp/d\varepsilon$ at zero temperature is determined by the density dependence
of the chemical potential:
\eq{\label{eq:vs2}
v_s^2 = \frac{n_B}{\mu_B} \, \frac{d \mu_B}{d n_B}~.
}
The quarkyonic transition typically predicts the existence of the peak structure in the $n_B$-dependence of this quantity~\cite{McLerran:2018hbz}. In the excluded volume quarkyonic matter model, this quantity exhibits singular behavior at the onset of quark appearance, as shown in the upper right panel of Fig.~\ref{fig:EoS}.
As first explored in Ref.~\cite{Jeong:2019lhv}, this behavior is caused by a sudden appearance of the Fermi sea of quarks and a rapid rise of the quark density of states near zero momentum $k_Q \approx 0$.
Obtaining a physically acceptable behavior of the speed of sound thus requires modifications to the quark sector of the model.
For instance, introducing an infrared regulator~\cite{Jeong:2019lhv}, corresponding to the multiplier $\frac{\sqrt{k^2+\Lambda^2}}{k}$ in the quark density of states~[the integrands in Eqs.~\eqref{eq:nQQM} and \eqref{eq:eQQM}], allows one to obtain a reasonable behavior of the speed of sound~(dotted blue line in Fig.~\ref{fig:EoS}).

In baryquark matter, the quarks appear on the Fermi shell rather than the Fermi sea; therefore, their density of states is vanishing at low momenta, $k_Q \approx 0$.
The associated behavior of the speed of sound is thus physically acceptable without the need to introduce any infrared regulators~(red line in Fig.~\ref{fig:EoS}).\footnote{We checked that introducing the infrared regulator $\sqrt{k^2+\Lambda^2}/k$ into baryquark matter lowers the energy per baryon at fixed density similar to quarkyonic matter, albeit in less dramatic fashion. Thus, baryquark matter stays energetically favored also in the presence of the regulator.}
Like in the various quarkyonic matter constructions, the speed of sound exhibits non-monotonic behavior and exceeds the conformal limit of $v_s^2 = 1/3$, although a pronounced peak is not observed.
This behavior is due to an early appearance of quarks in baryquark matter, which tames the rapid rise of the speed of sound due to the hard-core repulsion.

The early appearance of quarks in baryquark matter may be an artifact of neglecting attractive nuclear interactions relevant near the normal nuclear density.
To see this schematically, we apply a mean-field approach to incorporate the effect of attractive interactions, which implies a density-dependent contribution to nucleon energy density
\eq{\label{eq:attr}
\varepsilon_N \to \varepsilon_N - a \, n_N^2.
}
This model of nuclear interactions incorporating both the excluded volume and mean-field attraction corresponds to the quantum van der Waals equation of Ref.~\cite{Vovchenko:2015vxa}.

The dashed-dotted red lines in Fig.~\ref{fig:EoS} correspond to baryquark matter calculations incorporating attractive nuclear interactions with $a n_0 = 0.1 m_N$.
The presence of an attractive mean field lowers the energy density of nucleons at fixed $n_B$ per Eq.~\eqref{eq:attr} and thus disfavors the appearance of quarks at low densities.
For the present choice of parameters, the onset of quarks occurs at $n_B \simeq 3.4 \rho_0$.
This delayed appearance of quarks makes the equation of state stiffer and the peak in the sound velocity more pronounced.
Although in this simple model, $v_s^2$ exhibits a jump at the quark onset density~(dash-dotted line in Fig.~\ref{fig:EoS}), corresponding to a second-order phase transition, its behavior remains causal in the whole range of baryon densities.

The speed of sound behavior is closely related to the measure of conformality -- the trace anomaly, $\Delta = \frac{1}{3} - \frac{p}{\varepsilon}$. 
It has been suggested that the conformal limit, $\Delta = 0$, may be reached in the interior of neutron stars~\cite{Zeldovich:1961sbr}, and recent data-driven analyses indeed indicate such a possibility~\cite{Fujimoto:2022ohj,Marczenko:2022jhl}.
Our results indicate that this limit is reachable both in quarkyonic and baryquark matter if the equation of state is stiff enough, although this depends sensitively on the model parameters.
Interestingly, quarkyonic matter tends to yield a pronounced non-monotonic density dependence of $\Delta$, while in baryquark matter the approach to the conformal limit, $\Delta \to 0$, is smoother. 

\paragraph{\bf Discussion.}

Both quarkyonic and baryquark matter configurations are realizations of quark-hadron duality. 
In the quasi-particle picture of a baryon-quark mixture, baryquark matter turns out to be energetically favored.
To understand the reasons for this observation, let us compare the two scenarios in the following setup. 
Starting with a system at a given baryon density with nucleons only and neglecting nucleon interactions for a moment, we replace some of the nucleons with quarks in order to achieve a certain quark fraction, $n_Q/n_B$.
In case of baryquark matter, this simply entails replacing nucleons at the Fermi surface with quarks so that to leading order in $n_Q/n_B$ the energy does not change.
For quarkyonic matter, on the other hand, we not only need to replace the nucleons up to $k_F/N_c$ with quarks, but all the momentum states up to $k_F$ become unavailable for nucleons anymore, and thus these nucleons need to be moved to the Fermi surface, which costs additional energy.
This illustrates why it is energetically favorable to add deconfined quarks to the Fermi shell, as in baryquark matter, as opposed to the Fermi sea.
Nucleon interactions do not change this picture qualitatively, because to leading order they depend only on the nucleon density, which is the same for both scenarios.
Our calculations have been performed for $N_c = 3$ case, but the above considerations also apply to other values of $N_c$, including the large $N_c$ limit.

Conceptually, however, quarkyonic matter is arguably more appealing.
Consider the long-wavelength quark interactions, which one would typically associate with confinement. 
These interactions are Pauli-blocked in the Fermi sea but permitted on the Fermi surface.
One would thus identify the states in the Fermi shell with confined baryons rather than free quarks, as in quarkyonic matter.

The present results, which disfavor quarkyonic-like configurations as as the preferred form of dense QCD matter, were obtained in the quasiparticle picture of quarks and nucleons, with no internal motion of the confined quarks.
It is possible that this picture may be too naive to draw firm conclusions. 
Nevertheless, the results indicate that the presently explored realizations of quarkyonic matter cannot be regarded as fully consistent descriptions and would require modifications.
Indeed, any attempt to dynamically generate the quarkyonic-like momentum shell structure in a quasi-particle setup, for instance, via transport simulations, will inevitably end up in the more energetically preferred baryquark matter, if not in some other, more exotic configuration. 
One possibility is the introduction of momentum dependence into nuclear interactions -- a well known phenomenon (see e.g. \cite{Gale:1987zz}).
In order to disfavor baryquark matter, where on the average nucleons have smaller momenta than in quarkyonic matter, one would need an interaction which decreases with momentum, contrary to what is observed from proton-nucleus scattering experiments~\cite{Hama:1990vr}.
However, it cannot be ruled out that at densities above nuclear matter density the momentum-dependent part of interactions becomes attractive. For example recent nuclear matter ab initio calculations~\cite{Carbone:2019pkr}, albeit at finite temperature, find an increased effective mass possibly indicating an attractive interaction contribution due to momentum dependence.
Other extensions may include the introduction of quark interactions, perhaps by introducing a bag constant or in a parametric way~\cite{Fraga:2013qra}, as well as different choice of an infrared regulator.
Ultimately, the problem could be tackled by developing a framework where no momentum shell structure is imposed artificially but where the energetically favored configuration would emerge dynamically.

It is interesting to explore further the properties of the baryquark matter concept introduced here.
From a practical point of view, the model reasonably realizes the equation of state with a quark-hadron transition. Further refinements, for instance, to yield a more realistic description of the known nuclear matter properties at densities around the saturation, are certainly possible.
Another relevant application is isospin asymmetric matter and the associated neutron star phenomenology.
As in the case of quarkyonic matter, such an extension will involve dealing with multiple Fermi surfaces~(protons and neutrons) and the associated new phenomena and issues.
Finally, one can explore the consequences of the quark Fermi surface in baryquark matter.
This leads to the number of states at the Fermi surface being larger by a factor of $N_c$ compared to quarkyonic matter. Therefore, the heat capacity at low temperature for baryquark should be equally enhanced by a factor of $N_c$. 
Transport properties can also be expected to differ significantly between quarkyonic and baryquark scenarios.
In quarkyonic matter, one might expect transport properties to be similar to that of nuclear matter, given that the excitations near the Fermi surface are baryonic. In baryquark matter, on the other hand, the transport properties are more likely to resemble that of quark matter.

\paragraph{\bf Summary.}

We studied two opposite scenarios for the realization of the Pauli exclusion principle for an isospin symmetric mixture of quarks and nucleons at zero temperature: 
(i) a Fermi sea of quarks surrounded by a shell of baryons -- the quarkyonic matter, 
and 
(ii) a Fermi sea of nucleons surrounded by a shell of quarks which we defined as baryquark matter.
The 
energy minimization procedure
indicates that the baryquark matter is always energetically favored over the quarkyonic matter in this picture.
The resulting equations of state are qualitatively similar, describing the transition between baryon- and quark-dominated phases of matter.
In contrast to quarkyonic matter, however, the baryquark configuration yields a physically acceptable behavior of the speed of sound without the need to introduce an infrared regulator.
In order to retain the theoretically more appealing quarkyonic matter as the preferred form of dense QCD matter will thus require modifications to the existing dynamical generation mechanisms, such as, for example, the introduction of momentum-dependent nuclear interactions.

\begin{acknowledgments}

\emph{Acknowledgments.} 
We thank Yuki Fujimoto, Larry McLerran, Roman Poberezhnyuk, and Sanjay Reddy for fruitful discussions.
This work received support through the U.S. Department of Energy,
Office of Science, Office of Nuclear Physics, under contract numbers 
DE-AC02-05CH11231231 and DE-FG02-00ER41132.

\end{acknowledgments}

\bibliography{baryquark.bib}

\end{document}